\documentclass[twocolumn,superscriptaddress,floatfix,prl]{revtex4-1}
\usepackage{graphicx,amsfonts,amssymb,amsmath,hyperref,hypcap,enumerate}
\usepackage{color}
\usepackage{cleveref}

\newcommand{\bs}{\boldsymbol{s}}

\newcommand{\bA}{\boldsymbol{A}}
\newcommand{\BB}{\boldsymbol{B}}

\newcommand{\qq}{\boldsymbol{q}}

\newcommand{\RR}{\boldsymbol{R}}
\newcommand{\rr}{\boldsymbol{r}}

\newcommand{\kk}{\boldsymbol{k}}

\newcommand{\bb}{\boldsymbol{b}}

\newcommand{\nn}{\boldsymbol{n}}

\begin{document}
\title{Identification of superconducting pairing symmetry in twisted bilayer graphene using in-plane magnetic field and strain}

\author{Fengcheng Wu}
%\email{wufcheng@umd.edu}
\affiliation{Condensed Matter Theory Center and Joint Quantum Institute, Department of Physics, University of Maryland, College Park, Maryland 20742, USA}
\author{Sankar Das Sarma}
\affiliation{Condensed Matter Theory Center and Joint Quantum Institute, Department of Physics, University of Maryland, College Park, Maryland 20742, USA}

\date{\today}

\begin{abstract}
		We show how the pairing symmetry of superconducting states in twisted bilayer graphene can be experimentally identified by theoretically studying effects of externally applied in-plane magnetic field and strain. In the low field regime, superconducting critical temperature $T_c$ is suppressed by in-plane magnetic field $\BB_{\parallel}$ in singlet channels, but is  enhanced by weak $\BB_{\parallel}$ in triplet channels, providing an important distinction. The in-plane angular dependence of the critical $\BB_{\parallel, c}$ has a six-fold rotational symmetry, which is broken when strain is present. We show that anisotropy in $\BB_{\parallel, c}$ generated by strain can be similar for $s$- and $d$-wave channels in moir\'e superlattices. The $d$-wave state is pinned to be nematic by strain and consequently gapless, which is  distinguishable from the fully gapped $s$-wave state by scanning tunneling measurements.
\end{abstract}

\maketitle

{\it Introduction.---} The groundbreaking discovery \cite{Cao2018Super,Cao2018Magnetic} of correlated insulating  and superconducting (SC) states in twisted bilayer graphene (TBG)  has opened  the door to study many-body physics using versatile moir\'e bilayers. These initial findings have  been verified and expanded \cite{Dean2018tuning,kerelsky2018magic,choi2019imaging,sharpe2019emergent,MIT2018_rho,Columbia2018_rho,codecido2019correlated,lu2019superconductors,tomarken2019electronic}. New experimental developments include observations of giant linear-in-temperature resistivity in  the metallic regime  \cite{MIT2018_rho,Columbia2018_rho} and anomalous Hall effect at certain filling factors \cite{sharpe2019emergent,lu2019superconductors}. These remarkable discoveries have stimulated a large number of theoretical studies \cite{Balents2018,Senthil2018,Koshino2018, Kang2018,  Dodaro12018, Padhi2018, Guo2018,Fidrysiak2018,  Kennes2018strong,  Liu2018chiral,Isobe2018, You2018, Tang2019, rademaker2018charge,PALee2018, guinea2018electrostatic,Carr2018, Thomson2018, gonzalez2018kohn, Lin2018,Lado2018,Vishwanath2018origin, Ahn2018failure,Bernevig2018Topology, hejazi2018multiple, liu2018complete, sherkunov2018novel,Venderbos2018,KoziiNematic2019,xie2018nature,bultinck2019anomalous, Wu2018phonon, wu2019phonon,Heikkila2018, Lian2018twisted, choi2018electron} spanning almost the full gamut of solid-state physics from single-particle band structure theory to many-body theory. A key question in this context is the nature of the superconducting pairing symmetry, particularly whether any non-$s$-wave exotic pairing plays a role in TBG.

In this Letter, we focus on TBG SC states, which can be classified into $s$, $p$, $d$ and $f$ pairing channels based on the $D_6$ point group symmetry and spin SU(2) symmetry. All these SC channels have been theoretically proposed for TBG in the current literature.
The conventional $s$-wave pairing can arise from the enhanced electron-phonon interaction in the TBG flat bands  \cite{Wu2018phonon, wu2019phonon,Heikkila2018, Lian2018twisted, choi2018electron}. 
The unconventional pairing with $p$, $d$ or $f$ wave symmetry can be induced by electron-electron Coulomb repulsion \cite{Liu2018chiral,Isobe2018,You2018,Tang2019}, but can also be mediated by electron-phonon interaction  \cite{Wu2018phonon,wu2019phonon} due to TBG band symmetries, e.g., sublattice pseudospin chirality and valley symmetry. 

An important question is how the pairing symmetry in the TBG SC states can be experimentally identified.  We address this question by theoretically studying the response of SC states in each pairing channel  to in-plane magnetic field $\BB_{\parallel}$ and strain. We show that the dependence of SC critical temperature $T_c$ on  $\BB_{\parallel}$ in the low field regime distinguishes spin singlet ($s$, $d$) from triplet ($p$, $f$) pairings. In particular, $T_c$ is suppressed by $\BB_{\parallel}$ in singlet channels, but is {\it enhanced} by weak $\BB_{\parallel}$ in triplet channels. When compared to the first experiment\cite{Cao2018Super}, our theory indicates that the TBG SC states  has spin singlet pairing. For the singlet channels, $s$-wave state is fully gapped, while the $d$-wave state in the presence of strain is nematic and gapless. Therefore, $s$- and $d$-wave states can be distinguished by scanning tunneling gap measurement in the presence of applied weak strain.

{\it Moir\'e Hamiltonian.---}
The single-particle TBG physics  with small twist angle $\theta$ is described by the continuum moir\'e Hamiltonian \cite{Bistritzer2011},
which is independent of spin and is given in valley $\tau K$ by
\begin{equation}
\mathcal{H}_{\tau}=\begin{pmatrix}
h_{\tau \mathfrak{b}}(\kk) & T_{\tau }(\rr) \\
T^{\dagger}_{\tau}(\rr) & h_{\tau \mathfrak{t}}(\kk)
\end{pmatrix},
\label{Hmoire}
\end{equation}
where $h_{\tau \ell}$ is the Dirac Hamiltonian for layer $\ell$, and $T_{\tau }$ is the interlayer tunneling that varies spatially with the moir\'e period. Both $h_{\tau \ell}$ and $T_{\tau }$ can be specified using Pauli matrices $\sigma_{x, y, z}$ in the sublattice space as follows
\begin{equation}
h_{\tau \ell}(\kk) = \hbar v_F 
e^{-i\tau \ell \frac{\theta}{4} \sigma_z }[(\kk-\tau \boldsymbol{\kappa}_{\ell})\cdot (\tau \sigma_x, \sigma_y)]e^{+i\tau \ell \frac{\theta}{4} \sigma_z},
\end{equation}
\begin{equation}
\begin{aligned}
&T_{\tau}(\rr)=
T_{\tau}^{(0)}
+e^{-i \tau \bb_+ \cdot \rr} T_{\tau}^{(+1)}
+e^{-i \tau \bb_- \cdot \rr} T_{\tau}^{(-1)},\\
&T_{\tau}^{(j)} = w_0 \sigma_0 + w_1 \Big(\cos\frac{2 j \pi}{3} \sigma_x+ \tau \sin \frac{2 j \pi}{3} \sigma_y \Big),
\end{aligned}
\end{equation}
where $\tau=\pm$ is the valley index,  $\ell$ is   $+1$ and $-1$, respectively, for bottom ($\mathfrak{b}$) and top ($\mathfrak{t}$) layers. $\bb_{\pm}$ are moir\'e reciprocal lattice vectors given by $[4\pi/(\sqrt{3} a_M)](\pm1/2, \sqrt{3}/2)$, and momentum $\boldsymbol{\kappa}_{\ell}$ is equal to $[4\pi/(3 a_M)](-\sqrt{3}/2, -\ell/2)$ which connects the  center ($\gamma$) and one  corner of the moir\'e Brillouin zone. $a_M$ is the moir\'e period approximated by $a_0/\theta$, where $a_0$ is the monolayer graphene lattice constant. We use the same parameter values for $(v_F, w_0, w_1)$ as in Ref.~\cite{wu2019phonon}
%We take the bare Dirac velocity $v_F$ to be $10^6$ m/s, and the two tunneling parameters $w_0$ and $w_1$ to be 90 meV and 117 meV, respectively\cite{Jung2014}. Here $w_0$ and $w_1$ are different  due to the layer corrugation effect \cite{Koshino2018}.

We now discuss how the moir\'e Hamiltonian in Eq.~(\ref{Hmoire}) is modified by an in-plane magnetic field $\BB_{\parallel}$ and a strain field. The field $\BB_{\parallel}$ generates a layer-dependent gauge field $\bA_{\ell} = -\ell d_z (\BB_{\parallel} \times \hat{z})/2$, where $d_z$ is the interlayer distance. The momentum $\kk$ in $h_{\tau \ell}$ is then replaced by $\kk + e_0 \bA_{\ell}/\hbar $, where $e_0$ is the elementary charge. Besides this orbital effect, the magnetic field also leads to the Zeeman spin splitting term $\mu_B \bs \cdot \BB_{\parallel} $, where $\mu_B$ is the Bohr magneton and $\bs$ are spin Pauli matrices. %and a $g$-factor of 2 is used.

To study strain effect, we consider principal strains $\epsilon_1$ and $\epsilon_2$ respectively along two orthogonal directions $\nn_1=(\cos \phi, \sin \phi)$ and $\nn_2=(-\sin \phi, \cos \phi)$ such that  a vector parallel to $\nn_i$ is rescaled by a factor of $1+\epsilon_i$. This strain then transforms a generic vector $\RR$ to $(\hat{\mathcal{I}}_0+\hat{\mathcal{E}})\RR$, where $\hat{\mathcal{I}}_0 $ is identity matrix and $\hat{\mathcal{E}}$ is strain tensor defined as $\epsilon_1 \nn_1 \otimes \nn_1 + \epsilon_2 \nn_2 \otimes \nn_2.$
%\begin{equation}
%\begin{aligned}
%\hat{\mathcal{E}} \equiv
%\begin{pmatrix}
%\mathcal{E}_{xx} & \mathcal{E}_{xy} \\
%\mathcal{E}_{xy}  & \mathcal{E}_{yy}
%\end{pmatrix}=\epsilon_1 \nn_1^{T} \nn_1 + %\epsilon_2 \nn_2^{T} \nn_2.
%\end{aligned}
%\end{equation}
%Here $\mathcal{E}_{xx}= \epsilon_1 \cos^2 \phi + \epsilon_2 \sin^2 \phi$, $\mathcal{E}_{yy}= \epsilon_1 \sin^2 \phi + \epsilon_2 \cos^2 \phi$ and $\mathcal{E}_{xy}= (\epsilon_1-\epsilon_2)\sin \phi \cos \phi$.

We consider homostrain with the same strain field applied to bottom and top graphene layers (Heterostrain can be studied similarly\cite{Bi2019Strain}). Strain modifies the Dirac Hamiltonian $h_{\tau \ell}$ by shifting the Dirac point \cite{CastroNeto2009} as captured by the following Hamiltonian,
\begin{equation}
h_{\tau \ell}^{(1)} =
(\epsilon_1-\epsilon_2)\frac{ \hbar v_F }{2}\frac{\partial \ln |t_0|}{\partial a_{\text{CC}}} 
\begin{pmatrix}
0 & e^{i \tau (2 \phi+\ell \theta)} \\
e^{-i\tau (2\phi+\ell \theta)}  & 0
\end{pmatrix},
\end{equation}
where $t_0$ and $a_{\text{CC}}$ are respectively the nearest-neighbor hopping parameter and distance in monolayer graphene. Another effect of strain is that the momentum space vector $\boldsymbol{\kappa}_\ell$ is transformed to  $(\hat{I}_0+\hat{\mathcal{E}})^{-1} \boldsymbol{\kappa}_\ell$, and the same transformation also applies to $\bb_{\pm}$.

We assume that SC states in TBG have one of the pairing symmetries ($s$, $p$, $d$, $f$), and study effects of  $\BB_{\parallel}$ and $\hat{\mathcal{E}}$ fields on each channel in the following. 

\begin{figure}[t!]
	\includegraphics[width=1\columnwidth]{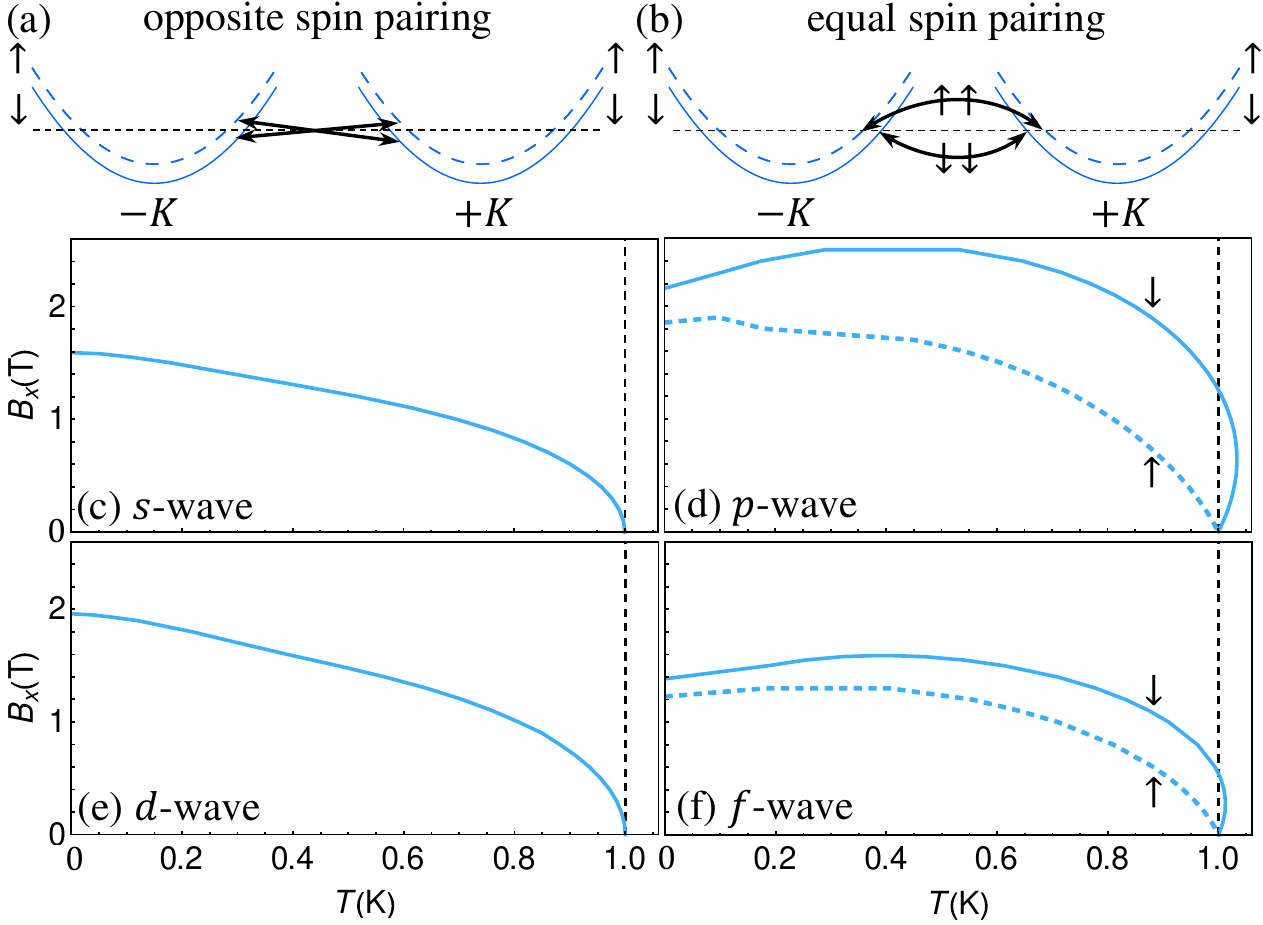}
	\caption{(a)-(b) Schematic illustration of the Zeeman effect on opposite spin pairing (singlet) and equal spin pairing (triplet).(c)-(f) In-plane critical magnetic field (along $\hat{x}$) as a function of temperature separately for the four channels. In the calculation,  twist angle $\theta$ is $1.1^{\circ}$, and the chemical potential $\mu$ is fixed to $\mu_0$, at which the moir\'e valence flat bands are half filled  when $\BB_{\parallel}=0$. The attractive interactions in each channel are adjusted so that $T_c=$1K at $\BB_{\parallel}=0$. In (d) and (f), solid and dashed lines respectively represent results in spin $\uparrow$ and $\downarrow$ components.}
	\label{Fig:B_T}
\end{figure}

{\it $s$-wave pairing.---} The $s$-wave channel can be realized by intrasublattice spin-singlet pairing. An $s$-wave pairing Hamiltonian can be derived from local on-site attractive interactions \cite{Wu2018phonon}, and is given by
\begin{equation}
H_0 =- g_0  \sum_{\tau_i}^{'} \sum_{\sigma, \ell} \int d\rr 
\hat{\psi}^{\dagger}_{\tau_1 \sigma \ell \uparrow}
\hat{\psi}^{\dagger}_{\tau_2 \sigma \ell \downarrow}
\hat{\psi}_{\tau_3 \sigma \ell \downarrow}
\hat{\psi}_{\tau_4 \sigma \ell \uparrow},
\label{Hs}
\end{equation}
where $\tau_i$ are valley indices with the constraint $\tau_1+\tau_2= \tau_3+\tau_4$, and  $\sigma$ represents  sublattices $A$ and $B$.  We use $\uparrow$ ($\downarrow$) to represent spin parallel (antiparallel) to $\BB_{\parallel}$.  The $s$-wave pair amplitudes for intervalley, intrasublattice and opposite-spin pairing are
\begin{equation}
\Delta_{s, \ell}(\rr) = s \langle \hat{\psi}_{- \sigma \ell (-s)}(\rr)  \hat{\psi}_{+ \sigma \ell s}(\rr) \rangle =  \sum_{\bb} e^{i \bb \cdot \rr} \Delta_{s,\ell,\bb}
\end{equation}
where  $s$ and $-s$ in the subscript represent opposite spins, $s$ in the prefactor  is $+1$($-1$) for spin $\uparrow$($\downarrow$), and $\bb$ represents moir\'e reciprocal lattice vectors. We make the ansatz that the pair amplitudes $\Delta_{s, \ell}(\rr)$ have the moir\'e periodicity and therefore, can be parametrized using  harmonic expansion.  The linearized gap equation is then given by
\begin{equation}
\begin{aligned}
&\Delta_{s,\ell,\bb} =  \sum_{\bb'\ell'} \chi_{\bb' \ell'}^{\bb \ell}(s) (\Delta_{\uparrow,\ell',\bb'}+\Delta_{\downarrow,\ell',\bb'}) \\
&\chi_{\bb' \ell'}^{\bb \ell}(s)=  \frac{  g_0}{2 \mathcal{A}} \sum_{\qq,n_1,n_2}  \mathcal{K}_{s (-s)} O_{\bb, \ell}^* O_{\bb', \ell'}, 
\end{aligned}
\label{chi_s}
\end{equation}
where the overlap function $O_{\bb, \ell}$ and the kernel function $\mathcal{K}_{s s'} $ are respectively defined as
\begin{equation}
\begin{aligned}
&O_{\bb, \ell}= \langle u_{n_1}(\qq,\BB_{\parallel}) | u_{n_2}(\qq,-\BB_{\parallel}) \rangle_{\bb, \ell}, \\
&\resizebox{0.9\columnwidth}{!} 
{$\displaystyle
\mathcal{K}_{s s'} 
=\frac{1-n_F[\varepsilon_{n_1}(\qq, \BB_{\parallel})+s\varepsilon_{\text{Z}}]-n_F[\varepsilon_{n_2}(\qq, -\BB_{\parallel})+s'\varepsilon_{\text{Z}}]}{\varepsilon_{n_1}(\qq, \BB_{\parallel})+s\varepsilon_{\text{Z}}+\varepsilon_{n_2}(\qq, -\BB_{\parallel})+s'\varepsilon_{\text{Z}}-2\mu} 
$.}
\end{aligned}
\label{OK}
\end{equation}
In Eq.~(\ref{OK}), $\mu$ is the chemical potential and $n_F$  the Fermi-Dirac occupation function. $| u_{n}(\qq,\pm \BB_{\parallel}) \rangle$ is the wave function of the $n$th moir\'e band at momentum $\qq$ in valley $+K$ for one spin component and in the presence of in-plane magnetic field $\pm \BB_{\parallel}$; $\varepsilon_{n}(\qq, \pm \BB_{\parallel})$ is the corresponding energy without including the Zeeman energy, while the Zeeman energy is taken into account by $\varepsilon_{\text{Z}}=\mu_B|\BB_{\parallel}|$. The overlap function $\langle ... \rangle_{\bb, \ell}$ represents the layer-resolved matrix element of the plane-wave operator $\exp(i \bb \cdot \rr )$. The SC critical temperature $T_c$ and critical $\BB_{\parallel}$ are obtained by solving Eq.~(\ref{chi_s}).

%We calculate $\chi$ by including momenta $\bb$ up to the third moir\'e reciprocal lattice vector shell, and by keeping only the nearly flat bands \cite{Wu2018phonon}. The superconducting critical temperature $T_c$ and critical $\BB_{\parallel}$ are obtained when Eq.~(\ref{chi_s}) is satisfied.

Spin $\uparrow$ and $\downarrow$ bands   are shifted by opposite energies due to the Zeeman effect, which leads to pair breaking for opposite spin pairing as in the $s$-wave channel [Fig.~\ref{Fig:B_T}(a)]. The orbital effect of $\BB_{\parallel}$ also results in depairing, because $\varepsilon_n(\qq, \BB_{\parallel}) \neq \varepsilon_n(\qq, -\BB_{\parallel})$.

%We note that $\Delta_{\uparrow, \ell}(\rr)=\Delta_{\downarrow, \ell}(\rr)$ when $\BB_{\parallel}$ is absent.

{\it $d$-wave pairing.---} The $d$-wave channel can be realized by {\it intersublattice} spin-singlet pairing, which can be mediated by intervalley optical phonons\cite{Wu2018phonon} as described by the following pairing Hamiltonian
\begin{equation}
\resizebox{0.88\columnwidth}{!} 
{$\displaystyle
H_d =- g_d \sum_{s, s', \ell} \sum_{\sigma \neq \sigma'}  \int d\rr  \hat{\psi}^{\dagger}_{+ \sigma \ell s}\hat{\psi}^{\dagger}_{- \sigma' \ell s'}\hat{\psi}_{+ \sigma \ell s'}\hat{\psi}_{- \sigma' \ell s},
$}
\label{Hd}
\end{equation}
where the main difference with Eq.~(\ref{Hs}) is in the sublattice dependence. The $d$-wave pair amplitudes are given by
\begin{equation}
\begin{aligned}
&\Delta_{s,\ell}^{(+)}(\rr) =s \langle \hat{\psi}_{- B \ell(-s)}(\rr)  \hat{\psi}_{+ A \ell s}(\rr) \rangle =  \sum_{\bb} e^{i \bb \cdot \rr} \Delta_{ s,\ell, \bb}^{(+)},\\
&\Delta_{s,\ell}^{(-)}(\rr) = s \langle \hat{\psi}_{- A \ell (-s)}(\rr)  \hat{\psi}_{+ B \ell s}(\rr) \rangle =  \sum_{\bb} e^{i \bb \cdot \rr} \Delta_{s,\ell,\bb}^{(-)},
\end{aligned}
\label{DeltaD}
\end{equation}
where  electrons in opposite valleys and different sublattices are paired. Here the orbital angular momentum of the $d$-wave pairing comes from the sublattice pseudospin chirality \cite{Wu2018phonon, wu2018topological}. The linearized gap equation is 
\begin{equation}
\begin{aligned}
&\Delta_{s,\ell,\bb}^{(L)} =  \sum_{\bb'\ell' L'} \chi_{\bb' \ell' L'}^{\bb \ell L}(s) \Delta_{(-s),\ell',\bb'}^{(L')}  \\
&\chi_{\bb' \ell' L'}^{\bb \ell L}(s)=  \frac{  g_d}{\mathcal{A}} \sum_{\qq,n_1,n_2}  \mathcal{K}_{s (-s)} \Omega_{\bb, \ell, L}^* \Omega_{\bb', \ell', L'}, \\
&\Omega_{\bb, \ell, L}= \langle u_{n_1}(\qq,\BB_{\parallel}) |\sigma_{L}| u_{n_2}(\qq,-\BB_{\parallel}) \rangle_{\bb, \ell},
\end{aligned}
\label{chi_d}
\end{equation}
where $L$ is the index ($\pm$) for the two chiral components in Eq.~(\ref{DeltaD}) and $\sigma_{L}$ is $(\sigma_x + i L \sigma_y)/2$. The kernel function $\mathcal{K}_{s(-s)}$ for the $d$-wave channel [Eq.~(\ref{chi_d})] is the same as the $s$-wave case [Eq.~(\ref{chi_s})], and therefore, the Zeeman effect also leads to depairing for $d$ wave. However, the overlap functions $O_{\bb,\ell}$ and $\Omega_{\bb,\ell,L}$ are different for the two channels, because of their different sublattice dependences.

When the $D_6$ symmetry is preserved (i.e., magnetic field $\BB_{\parallel}$ and strain are absent), the $d$-wave linearized gap equation leads to two degenerate states that can be classified as $d_{x^2-y^2}$ and $d_{xy}$ nematic states (alternatively, $d+i d$ and $d-i d$ chiral states). The two-component nature of the $d$-wave pairing is characterized by a nematic director  that can be chosen as
\begin{equation}
\boldsymbol{\eta} 
\equiv (\eta_x, \eta_y)
= \Big(\frac{\Delta_{\uparrow,\mathfrak{b}}^{(+)}-\Delta_{\uparrow,\mathfrak{b}}^{(-)}}{\sqrt{2} i},\frac{\Delta_{\uparrow,\mathfrak{b}}^{(+)}+\Delta_{\uparrow,\mathfrak{b}}^{(-)}}{\sqrt{2}} \Big)|_{\rr=0},
\end{equation}
where $\rr=0$ corresponds to the center of the $AA$ region in the moir\'e pattern. 
The $d_{x^2-y^2}$ and $d_{xy}$  states respectively correspond to $\boldsymbol{\eta} \propto (1,0) $ and $\boldsymbol{\eta} \propto (0,1) $, and their degeneracy at $T_c$ is lifted by magnetic and strain fields.
A  nematic order parameter \cite{Fu2014} can be further defined as
\begin{equation}
 \boldsymbol{N}=(|\eta_x|^2-|\eta_y|^2, \eta_x^*\eta_y+\eta_y^*\eta_x),
\end{equation}
which is gauge invariant and will be used to characterize the $d$-wave states in the following.

\begin{figure}[t!]
	\includegraphics[width=0.9\columnwidth]{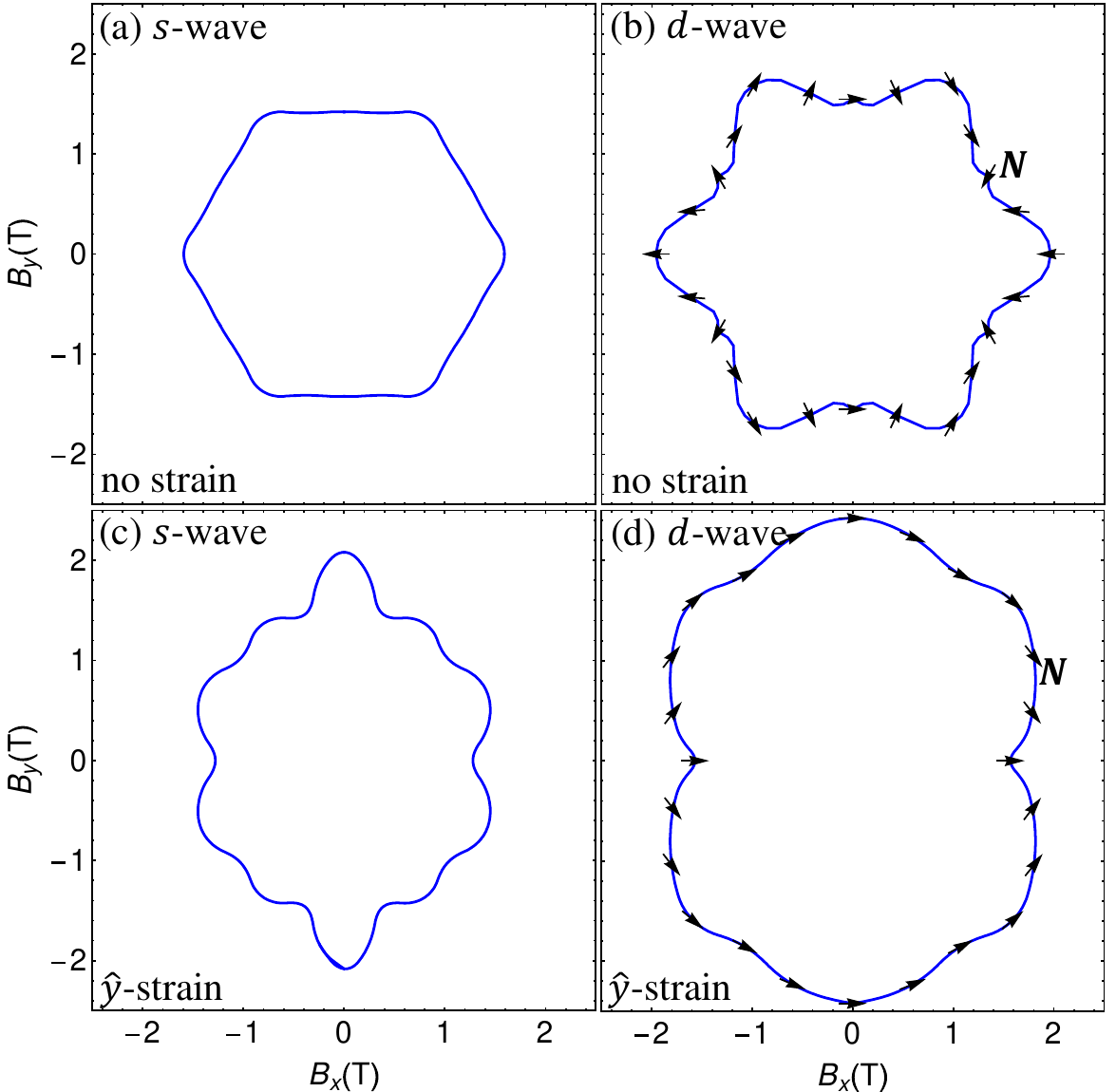}
	\caption{In-plane angular variation of $\BB_{\parallel, c}$ at zero temperature. (a) and (b) are respectively for $s$- and $d$-wave channels without strain. (c) and (d) are corresponding plots with a strain (0.4\%) applied along $\hat{y}$ direction. In (b) and (d), the black arrows indicate the nematic order parameter $\boldsymbol{N}$. In (d), $\boldsymbol{N}$ is primarily pinned to $+\hat{x}$ direction due to the strain. }
	\label{Fig:BxBy}
\end{figure}

{\it $f$-wave pairing.---}The $f$-wave channel can be realized by {\it intrasublattice} spin-triplet pairing, which can be mediated by acoustic phonons \cite{wu2019phonon}. An in-plane magnetic field $\BB_{\parallel}$ favors equal spin pairings in the $f$-wave channel. The equal-spin $f$-wave pairing Hamiltonian is given by
\begin{equation}
H_f =- g_f \sum_{\sigma,\ell}\sum_{s=\uparrow,  \downarrow}\int d\rr \hat{\psi}_{+ \sigma \ell s}^\dagger  \hat{\psi}_{- \sigma \ell s}^\dagger
\hat{\psi}_{- \sigma \ell s} \hat{\psi}_{+ \sigma \ell s}.
\label{Hf}
\end{equation}
The corresponding pair amplitude and linearized gap equation are 
\begin{equation}
\begin{aligned}
&\Gamma_{s, \ell}(\rr) = \langle \hat{\psi}_{- \sigma \ell s}(\rr)  \hat{\psi}_{+ \sigma \ell s}(\rr) \rangle =  \sum_{\bb} e^{i \bb \cdot \rr} \Gamma_{s,\ell,\bb},\\
&\Gamma_{s,\ell,\bb} =  \sum_{\bb'\ell'} \tilde{\chi}_{\bb' \ell'}^{\bb \ell}(s) \Gamma_{s,\ell',\bb'}, \\
&\tilde{\chi}_{\bb' \ell'}^{\bb \ell}(s)=  \frac{  g_f}{2 \mathcal{A}} \sum_{\qq,n_1,n_2}  \mathcal{K}_{s s} O_{\bb, \ell}^* O_{\bb', \ell'}, 
\end{aligned}
\label{chi_f}
\end{equation}
which are similar to the $s$-wave case except that electrons with the same spin are paired. The two spin components ($\uparrow$, $\downarrow$) have independent gap equations as shown in Eq.~(\ref{chi_f}), and therefore, separate $T_c$ values.  The Zeeman effect leads to different effective chemical potentials for the two independent spin components, which have important consequences as discussed below.

{\it $p$-wave pairing.---}The $p$-wave channel can be realized by {\it intersublattice} spin-triplet pairing, which can again be mediated by acoustic phonons \cite{wu2019phonon}. The equal-spin $p$-wave pairing Hamiltonian is given by
\begin{equation}
\resizebox{0.88\columnwidth}{!} 
{$\displaystyle
H_p =- g_p \sum_{s=\uparrow,\downarrow} \sum_{\sigma \neq \sigma',\ell} \int d\rr \hat{\psi}^{\dagger}_{+ \sigma \ell s}\hat{\psi}^{\dagger}_{- \sigma' \ell s}\hat{\psi}_{- \sigma' \ell s}\hat{\psi}_{+ \sigma \ell s} 
.
$}
\label{Hp}
\end{equation}
The analysis of $p$-wave channel is similar to the $d$-wave  except for the difference of equal spin pairing. The Zeeman effect plays the same role in $p$- and $f$-wave channels.

\begin{figure}[t!]
	\vspace{0 mm}
	\includegraphics[width=1\columnwidth]{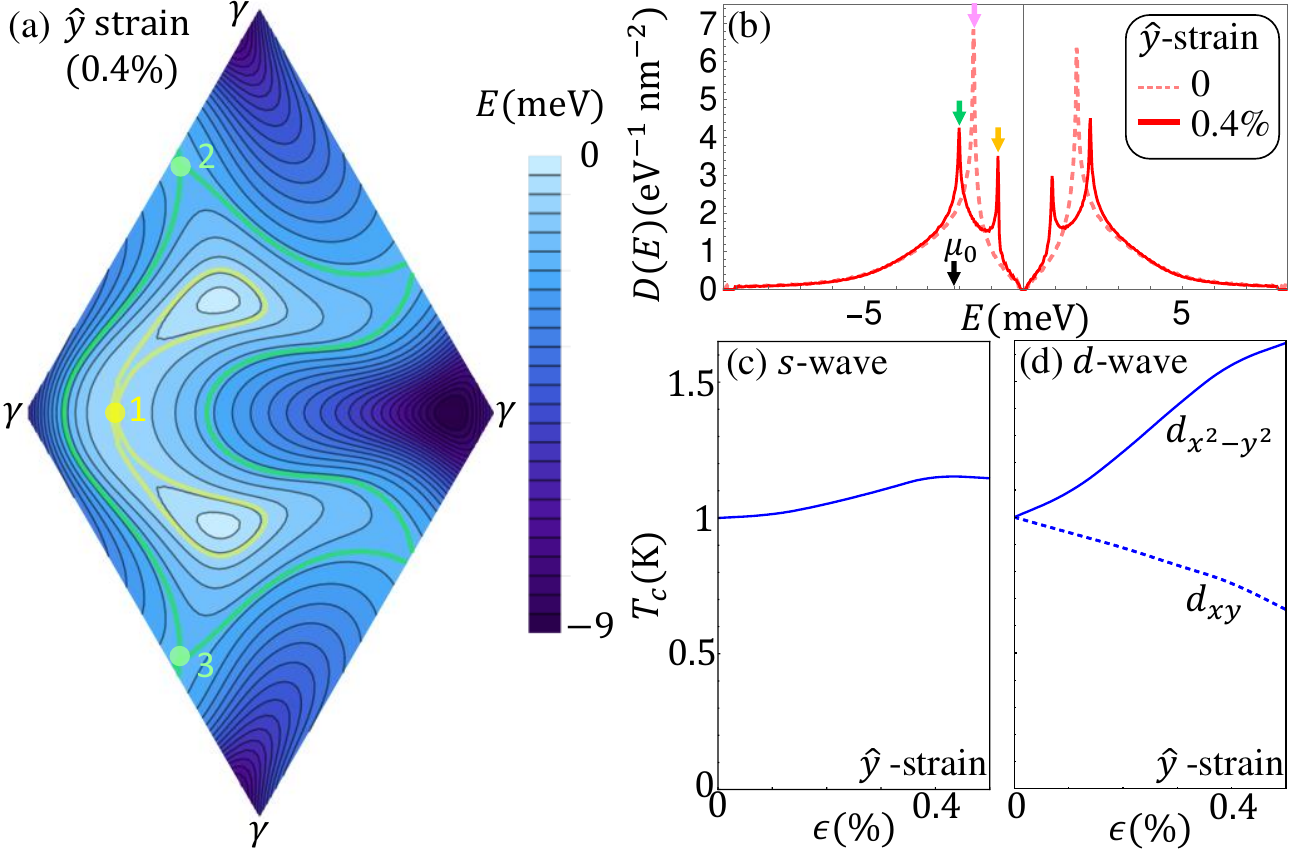}
	\vspace{-5 mm}
	\caption{(a) Energy contour plot for the moir\'e valence band in $+K$ valley with strain (0.4\%) applied along $\hat{y}$ direction. Dots 1, 2 and 3 represent saddle points, which would be related by $\hat{C}_3$ and degenerate if strain were absent. (b) DOS per spin and per valley with (solid line) and without (dashed line) strain. (c) and (d) $T_c$ as a function of strain $\epsilon$ along $\hat{y}$ direction. The chemical potential is fixed to $\mu_0$ as in Fig.~\ref{Fig:B_T}. In (d), solid and dashed lines respectively represent the temperature at which the largest and the second largest eigenvalues of the $d$-wave susceptibility reach 1. $T_c$ is given by the solid line, at which $d_{x^2-y^2}$ pairing is formed. Below the dashed line, $d_{xy}$ pairing could set in to form chiral $d$-wave state.  }
	\label{Fig:strain}
\end{figure}

{\it Critical $\BB_{\parallel}$.---}
We calculate the in-plane critical magnetic field as a function of temperature. Representative results in the absence of strain field are plotted in Figs.~(\ref{Fig:B_T})(c) to (\ref{Fig:B_T})(f) respectively for the four channels. When the field ($\BB_{\parallel}$) is weak, the dependence of $T_c$ on $\BB_{\parallel}$ shows a clear distinction between singlet and triplet pairing channels. In the singlet $s$- and $d$-wave channels, $T_c$ is suppressed by  $\BB_{\parallel}$, which can be described by a  quadratic function $T_c = T_c (0)(1-  |\BB_{\parallel}|^2/B_0^2$). However, in the triplet $p$- and $f$-wave channels, $T_c$ can actually be enhanced slightly by the weak $\BB_{\parallel}$ field. This is due to the Zeeman effect, which leads to different effective chemical potential $\mu_{\uparrow}=\mu-\varepsilon_{\text{Z}}$ and $\mu_{\downarrow}=\mu+\varepsilon_{\text{Z}}$ respectively for the two spin components. Quite generally, the density of states (DOS) is, respectively, enhanced and suppressed at the two effective chemical potentials ($\mu_{\uparrow},\mu_{\downarrow}$) compared to DOS at the original chemical potential $\mu$. Therefore, $T_c$ increases and decreases {\it linearly} with weak $\BB_{\parallel}$  field respectively in the two spin components   [Figs.~(\ref{Fig:B_T})(d) and (\ref{Fig:B_T})(f)]. 
It can be shown by examining the triplet pair susceptibility [Eq.~(\ref{chi_f})] that the leading orbital effect on $T_c$ is second order, while the Zeeman effect dominates in the low-field regime for equal spin pairing. \cite{SM}
When the field becomes strong enough, $T_c$ is suppressed in both spin components due to the orbital effect.

The experimental dependence of $T_c$ on $\BB_{\parallel}$ for the SC states in TBG \cite{Cao2018Super} can be well fitted by $T_c = T_c (0)(1-  |\BB_{\parallel}|^2/B_0^2$), which is consistent with spin singlet pairing.

The in-plane angular dependence of the critical $\BB_{\parallel,c}$ at zero temperature ($T=0$) is shown in Fig.~\ref{Fig:BxBy}(a) and \ref{Fig:BxBy}(b), respectively, for $s$- and $d$-wave channels. In both channels, the angular dependence of $\BB_{\parallel,c}$ is six-fold rotational symmetric, reflecting the symmetry of the underlying lattice. $\BB_{\parallel,c}$ has angular variation because of the magnetic orbital effect that is effectively magnified in the nearly flat bands close to magic angle. In the $d$-wave channel, $\BB_{\parallel}$ explicitly lifts the degeneracy between the two nematic states. For example, $d_{xy}$ and $d_{x^2-y^2}$ states are respectively favored for $\BB_{\parallel}$ along $\hat{x}$ and $\hat{y}$ directions, as illustrated in Fig.~\ref{Fig:BxBy}(b).

\begin{figure}[t!]
	\includegraphics[width=1\columnwidth]{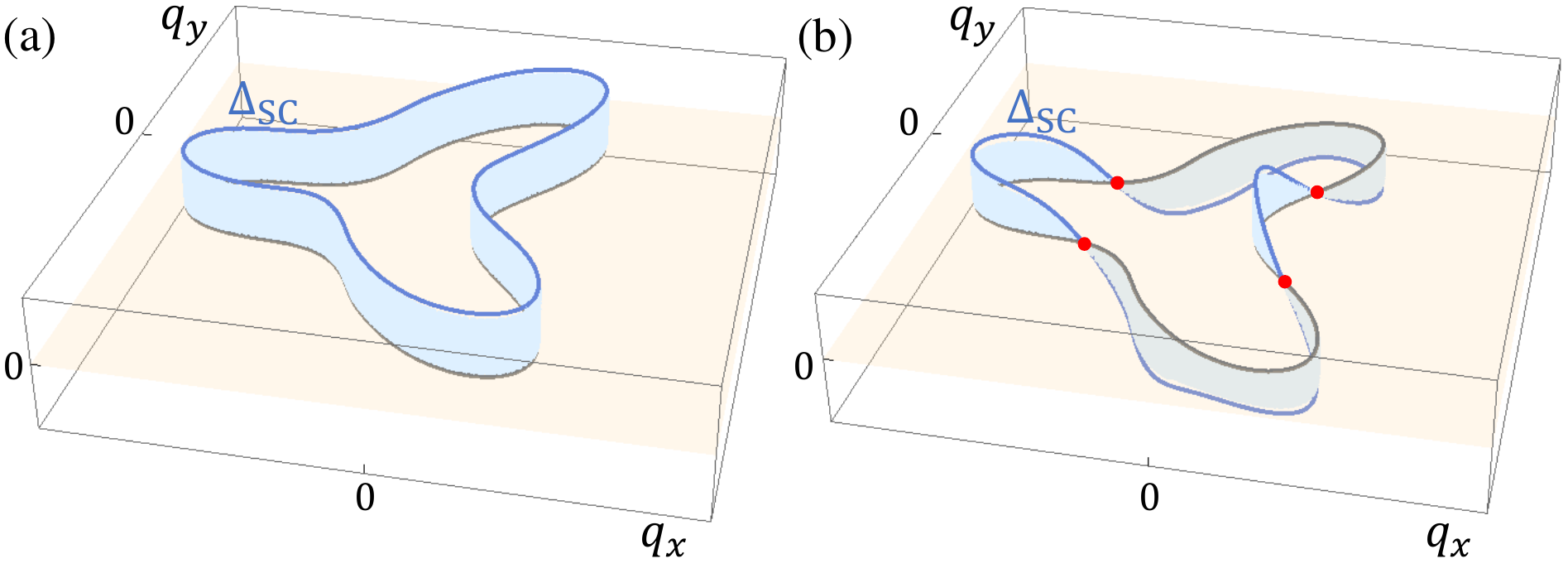}
	\caption{(a) $s$-wave SC gap function (blue line) along the Fermi surface (gray line). (b) Corresponding plot for $d_{x^2-y^2}$ state, which has point nodes as marked by the four red dots. }
	\label{Fig:gap}
\end{figure}

{\it Strain tunability.---}
%We turn to demonstrate how superconductivity in TBG can be tuned by strain. 
A representative moir\'e band structure in the presence of strain is depicted in Fig.~\ref{Fig:strain}(a). One important effect of strain is to break the three-fold rotational symmetry $\hat{C}_3$. As a manifestation, strain lifts the degeneracy among moir\'e band saddle points that would be related to each other by $\hat{C}_3$ if strain were absent. Therefore, a van Hove singularity in the DOS can be split into multiple van Hove singularities by  strain, as shown in Fig.~\ref{Fig:strain}(b). Consequently, $T_c$ is tunable by strain, as plotted in Figs.~\ref{Fig:strain}(c) and \ref{Fig:strain}(d) respectively for $s$- and $d$-wave channels. Besides this quantitative effect on $T_c$, strain also has a qualitative effect in $d$-wave channel by lifting the degeneracy between $d_{x^2-y^2}$ and $d_{xy}$ at $T_c$ [Fig.~\ref{Fig:strain}(d)].  Uniaxial strain pins the nematic order parameter $\boldsymbol{N}$ to a particular direction. For example, tensile strain applied along $\hat{x}$ ($\hat{y}$) favors $d_{xy}$ ($d_{x^2-y^2}$) nematic state, similar to the effect of in-plane magnetic field.\cite{SM}

In the presence of a fixed uniaxial strain, the angular variation of the in-plane critical magnetic field $\BB_{\parallel,c}$ is twofold  (instead of sixfold) rotational symmetric. The strain can lead to similar anisotropy in $\BB_{\parallel,c}$ 
for $s$- and $d$-wave channels [Figs.~\ref{Fig:BxBy}(c) and \ref{Fig:BxBy}(d)], because the nearly flat moir\'e bands near magic angle are sensitive to tiny strain. This raises the question how $s$- and $d$-wave channels can be distinguished experimentally. 
We propose that scanning tunneling microscopy (STM) study of the SC gap can be used for the distinction. In particular, the $s$-wave superconductivity is fully gapped [Fig.~\ref{Fig:gap}(a)]. In contrast, the nematic $d$-wave state, which is pinned by strain, is gapless with point nodes [Fig.~\ref{Fig:gap}(b)].

{\it Discussion.---} 
%Our theory shows that the dependence of SC critical temperature on weak in-plane magnetic field can be used to distinguish spin singlet and triplet pairings. Comparing our theory to the experiment in Ref.~\cite{Cao2018Super}, we conclude that the SC states in TBG has spin singlet pairing. 
Strain  has been found to be generally presented in TBG devices from STM imaging \cite{kerelsky2018magic,choi2019imaging}. Recent transport study \cite{MIT_nematic} shows that  in-plane critical magnetic field $\BB_{\parallel,c}$ for SC states in TBG has a two-fold anisotropy, which also indicates the presence of strain. This two-fold anisotropy in $\BB_{\parallel,c}$ might be a signature of nematic state, because SC states with a two-component order parameter such as $d$-wave channel have been argued to be more susceptible to strain compared to $s$-wave states in bulk materials \cite{Fu2014,Venderbos_Hc2}. However, we note that moir\'e band structure itself can be very sensitive to strain due to the narrow bandwidth, and our calculation indicates that strain can in principle lead to similar anisotropy in $\BB_{\parallel,c}$ for $s$- and $d$-wave channels. Therefore, we suggest that a tunneling measurement of SC gap could provide a more conclusive distinction between these two pairing states.

Evidences of superconductivity have recently been reported also in other  moir\'e systems, for example, $ABC$ trilayer graphene on boron nitride \cite{Chen2019SC} and twisted double bilayer graphene \cite{shen2019observation,liu2019spin,cao2019electric}. Our theory can be generalized to these related systems. In twisted double bilayer graphene, in-plane magnetic field has been found to enhance  $T_c$ in the low field regime \cite{liu2019spin}, which has been interpreted as an indication of spin triplet pairing \cite{liu2019spin,lee2019theory}; our theory supports this interpretation.

{\it Acknowledgment.---}This work is supported
by Laboratory for Physical Sciences.

\bibliographystyle{apsrev4-1}
\bibliography{refs}

%\clearpage
\begin{center}
	\textbf{Supplemental Material}
\end{center}
%\section{Supplemental Material}

\section{Symmetries}

In our continuum theory, TBG has $D_6$ point group symmetry, which is generated by a six-fold rotation $\hat{C}_{6z}$ around $\hat{z}$ axis, and twofold rotations $\hat{C}_{2x}$ and $\hat{C}_{2y}$, respectively, around $\hat{x}$ and $\hat{y}$ axes. Note that $\hat{C}_{6z}$ generates $\hat{C}_{3z} = (\hat{C}_{6z})^2$ and $\hat{C}_{2z} = (\hat{C}_{6z})^3$. All directions related to $\hat{x}$ or $\hat{y}$ by $\hat{C}_{6z}$ are also twofold rotation axes, which we refer to as in-plane twofold rotation axes. Electrons in TBG have spin SU(2) symmetry and spinless time-reversal symmetry $\hat{\mathcal{T}}$. The $\hat{C}_{2z} \hat{\mathcal{T}}$ enforces the Berry curvature to be zero and protects the Dirac cone band touching.

An applied in-plane magnetic field $\BB_{\parallel}$ breaks the $\hat{\mathcal{T}}$ symmetry as well as all rotational symmetries around $\hat{z}$ axis. However, $\BB_{\parallel}$ preserves $\hat{C}_{2z} \hat{\mathcal{T}}$, and therefore, Dirac cones remain gapless, although the position of Dirac points in momentum space changes with $\BB_{\parallel}$. 

Strain preserves $\hat{\mathcal{T}}$ symmetry and $\hat{C}_{2z}$ symmetry, but breaks  other rotational symmetries around $\hat{z}$ axis. Again, $\hat{C}_{2z} \hat{\mathcal{T}}$ symmetry is preserved and Dirac cones remain gapless with movable Dirac touching points.  When $\BB_{\parallel}$ or strain is along one of the in-plane twofold rotation axes, there is a two-fold rotational symmetry around this axis.

We denote the moir\'e band energy (without including Zeeman term) as $\varepsilon_{n,\tau}(\qq, \BB_{\parallel})$, where $n$ is the band index and $\tau$ is the valley index. Because time reversal operation flips $\tau$, $\qq$ and $\BB_{\parallel}$, we have the identity  $\varepsilon_{n,\tau}(\qq, \BB_{\parallel}) = \varepsilon_{n,-\tau}(-\qq, -\BB_{\parallel})$, which has been used to simplify the pair susceptibilities, for example, Eq.~(8) of the main text.

\section{orbital effect of magnetic field  on $T_c$}
We show that the leading orbital effect of in-plane magnetic field on $T_c$ is second order, while the Zeeman splitting has a dominant first order effect on $T_c$ for the spin triplet channel in the low field regime. 

For definiteness, we consider in-plane magnetic field $\BB_{\parallel}$ along $\hat{x}$ direction with $\BB_{\parallel} = B {\hat{x}}$. We use $\rightarrow$ and $\leftarrow$ to indicate spin pointing to $+\hat{x}$ and $-\hat{x}$ directions, respectively. The Zeeman splitting is denoted as $\varepsilon_x = \mu_B B$, where $\varepsilon_x$ can be positive or negative depending on $B$. For equal spin pairing, $\rightarrow$ and $\leftarrow$ spin components have independent linearized gap equations. To demonstrate the effect of $\BB_{\parallel}$, we consider $\rightarrow$ component in the $f$-wave channel, of which the pair susceptibility is
\begin{equation}
\begin{aligned}
&\chi_{\rightarrow}(\BB_{\parallel},\varepsilon_x)=\frac{g_f}{4\mathcal{A}}\sum_{\qq} \mathcal{K}_{\rightarrow} |O|^2,\\
&O= \langle u(\qq,\BB_{\parallel}) | u(\qq,-\BB_{\parallel}) \rangle, \\
&\mathcal{K}_{\rightarrow} 
	=\frac{1-n_F[\varepsilon(\qq, \BB_{\parallel})+\varepsilon_{x}]-n_F[\varepsilon(\qq, -\BB_{\parallel})+\varepsilon_{x}]}{\varepsilon(\qq, \BB_{\parallel})+\varepsilon(\qq, -\BB_{\parallel})-2(\mu-\varepsilon_{x})},
\end{aligned}
\label{chix}
\end{equation}
where we have retained only the moir\'e band that crosses the Fermi energy, and the $\bb=0$ component of the susceptibility. The bottom and top graphene layers are assumed to have the same pair amplitudes in Eq.~(\ref{chix}). These simplifications make the analysis more straightforward.

Equation (\ref{chix}) shows that $\chi_{\rightarrow}(\BB_{\parallel},\varepsilon_x) = \chi_{\rightarrow}(-\BB_{\parallel},\varepsilon_x)$, i.e, $\chi_{\rightarrow}$ remains the same when  the magnetic field changes sign only in the  orbital effect  but not in the Zeeman effect. Therefore, the orbital effect of magnetic field on $T_c$ is at least second order. The Zeeman splitting effectively shifts the chemical potential $\mu$ to $\mu-\epsilon_x$, which results in a first-order correction to $T_c$ as given by
\begin{equation}
\begin{aligned}
k_B T_{\rightarrow, c} & \approx \Lambda \exp[- \frac{1}{\tilde{g}_f D(\mu-\epsilon_x)}]\\
& \approx \Lambda \exp[- \frac{1}{\tilde{g}_f D(\mu)}] [1-\frac{D'(\mu)}{\tilde{g}_f D^2(\mu)) } \mu_B B ],
\end{aligned}
\end{equation}
where $\Lambda$ is an energy cutoff, $\tilde{g}_f=g_f/4$, and $D(\mu)$ is the DOS. The variation of DOS with $\mu$ leads to a linear $B$ correction to $T_c$ in the triplet channel.

It can be shown in a similar way that both  orbital and Zeeman effects lead to  second-order corrections to $T_c$ for spin singlet channels.

\section{Ginzburg-Landau theory for $d$-wave channel}

The $d$-wave state has a two-component pairing order parameter. As demonstrated in the main text, the nematic director $\boldsymbol{\eta}$ can be steered by in-plane magnetic field and strain field, which is captured by the phenomenological Ginzburg-Landau free energy
\begin{equation}
\begin{aligned}
\mathcal{F}_d = &\alpha [T-(T_c(0)-\beta |\BB_{\parallel}|^2)](|\eta_x|^2+|\eta_y|^2)\\
&+[\lambda_1(B_x^2-B_y^2)+\lambda_2(\mathcal{E}_{xx}-\mathcal{E}_{yy}) ](|\eta_x|^2-|\eta_y|^2)\\
&+(2\lambda_1 B_x B_y + 2 \lambda_2 \mathcal{E}_{xy} )(\eta_x^* \eta_y + \eta_y^* \eta_x),
\end{aligned}
\label{Fd}
\end{equation}
where we keep leading order terms, in particular,  second order terms in   $\boldsymbol{\eta}$ and $\BB_{\parallel}$, and first order terms in strain. $\mathcal{E}_{xx}$ and
$\mathcal{E}_{yy}$ are diagonal terms in the symmetric strain tensor $\hat{\mathcal{E}}$, and $\mathcal{E}_{xy}$ is the off-diagonal term.
The numerical results shown in Figs. 2 and 3 of the main text correspond to $\lambda_1>0$ and $\lambda_2>0$ such that $d_{xy}$ ($d_{x^2-y^2}$) state is preferred when $\BB_{\parallel}$ or tensile strain is along $\hat{x}$ ($\hat{y}$). Because the free energy $\mathcal{F}_d$ in Eq.~(\ref{Fd}) only includes terms up to second order, it does not fully capture the angular variation of  the in-plane critical magnetic field $\BB_{\parallel, c}$ shown in Fig. 2(b) of the main text.

%\bibliographystyle{apsrev4-1}
%\bibliography{refs}

\end{document}